\documentclass[prl,twocolumn,showpacs,superscriptaddress]{revtex4-1}
\usepackage{graphicx,amssymb,amsmath,psfrag,color,hyperref}

\definecolor{darkgreen}{rgb}{0,0.5,0}
\definecolor{purple}{rgb}{0.35,0,0.35}
\definecolor{orange}{rgb}{1,0.5,0}
\definecolor{darkred}{rgb}{.7,0,0}
\definecolor{darkblue}{rgb}{0,0,.3}
\definecolor{grey}{rgb}{.6,.6,.6}
\definecolor{dimgreen}{rgb}{0.2,0.6,0.1}

\newcommand{\jav}[1]{{#1}}

\begin{document}
\definecolor{darkgreen}{rgb}{0,0.5,0}

\title{Quantum quench in PT-symmetric Luttinger liquid}

\author{Bal\'azs D\'ora}
\email{dora@eik.bme.hu}
\affiliation{Department of Theoretical Physics and MTA-BME Lend\"ulet Topology and Correlation Research Group,
Budapest University of Technology and Economics, 1521 Budapest, Hungary}
\author{C\u at\u alin Pa\c scu Moca}
\affiliation{Department of Theoretical Physics and
BME-MTA  Exotic  Quantum  Phases  Research Group,   Budapest  University  of  Technology  and  Economics,  1521  Budapest,  Hungary}
\affiliation{Department  of  Physics,  University  of  Oradea,  410087,  Oradea,  Romania}

\date{\today}

\begin{abstract}
A Luttinger liquid (LL) describes low energy excitations of many interacting one dimensional systems, and exhibits
universal response both in and out of equilibrium.
 We analyze its behaviour in the non-hermitian realm after 
quantum quenching to a PT-symmetric LL by focusing on the fermionic single particle density matrix.
For short times, we demonstrate the emergence of unique phenomena, characteristic to non-hermitian systems, that
correlations propagate faster than the conventional maximal speed, known as the Lieb-Robinson bound.
These emergent supersonic modes travel with velocities that are multiples of the conventional light-cone velocity.
This behaviour is argued to be generic for correlators in non-hermitian systems.
In the long time limit, we find typical LL behaviour,  extending the LL universality to the non-equilibrium non-hermitian case.
Our analytical results are benchmarked numerically and indicate that the dispersal of quantum information is much faster in non-hermitian systems.
\end{abstract}
\maketitle

\paragraph{Introduction.}
Non-hermitian quantum mechanics has provided us with a plethora of interesting phenomena, investigated both theoretically and experimentally.
These include spontaneous PT-symmetry breaking, non-unitary dynamics, encircling and manipulating exceptional points, unidirectional invisibility, 
complex Bloch oscillations and even topological 
effects\cite{ptreview,rotter,bender98,berry2004,ganainy,xiao,gao2015,zhou18,tonylee,zeuner,yao18}, to
mention a few. 
However, the common theme behind these studies is the underlying effective single particle picture, 
while excursions to the quantum many-body realm are scarce\cite{ashida}.

By now, hermitian quantum many-body physics in one spatial dimension is well understood thanks to the 
available analytical and numerical methods\cite{giamarchi,nersesyan}, both in and out of equilibrium.
In particular, many of these systems realize Luttinger liquids (LLs), including bosonic, 
fermionic, spin etc. models, irrespective of their statistics and microscopic details\cite{giamarchi,nersesyan,cazalillarmp}.
Therein, the Fermi-liquid description breaks down and the elementary excitations are bosonic collective modes. 
This fractionalization is manifested by the universal non-integer power law decays in almost all correlation functions\cite{giamarchi}.
Furthermore, when taken out of equilibrium,  for example following a quantum quench, 
the evolution of such hermitian systems is always unitary. The changes of system parameters cause the emission of quasiparticles carrying 
correlations that propagate across the system with a certain velocity
 whose maximum value is given by the Lieb-Robinson bound\cite{liebrobinson}. The existence of a maximum  
 speed implies a light-cone spreading of correlations\cite{Calabrese_2005}  with only subsonic mode velocities. 

Therefore, it comes as a natural question  whether any of the LL universality and light-cone structure survive under non-hermitian conditions.
This motivated us to explore the fate of LLs after a quantum quench to a PT-symmetric non-hermitian system. 
We find that the long time limit exhibits
LL behaviour with non-integer power law decays, thus extending LL universality to the non-equilibrium non-hermitian realm.

The short time behaviour, on the other hand, differs drastically from that in the hermitian realm\cite{cazalillaprl}. 
On top of the usual light-cone\cite{liebrobinson}, 
new supersonic modes appear and travel with velocities that are multiples of
 the light-cone velocity. 
The origin of the supersonic modes is related to an effective long range Hamiltonian, governing the time evolution, 
for which such supersonic modes might be expected\cite{eisert13}.
We argue that this emergent phenomenon is characteristic to all 
correlation functions in non-hermitian systems.
These findings are tested numerically on a non-hermitian variant of the XXZ Heisenberg model. Its density correlation function reveals three distinct light-cones, in perfect agreement with bosonization.

\paragraph{PT-symmetric LL.}
The non-hermitian LL Hamiltonian, we study, is given by
\begin{equation}
H=\sum_{q\neq 0} v|q|b_q^\dagger b_q
+\frac{ig_2|q|\Theta(t)}{2}\left[b_qb_{-q}+b_q^+b_{-q}^+\right],
\label{hamilton}
\end{equation}
with $v$ being the bare "sound velocity", 
and $b_q^\dagger$ the creation operator of a bosonic density wave. 
The interaction $g_2$ is changed 
 from zero to a nonzero value at $t=0$.
Although the Hamiltonian is non-hermitian, its spectrum remains real\cite{garcia} as $\omega_q=\tilde v |q|$ in the presence of imaginary interaction, with renormalized velocity $\tilde v=\sqrt{v^2+g_2^2}$.
This falls into the category of PT-symmetric non-hermitian systems\cite{ptreview,rotter,bender98,berry2004}: Eq. \eqref{hamilton} satisfies 
the antiunitary (generalized PT-) symmetry\cite{garcia} as the combination of time reversal, $i\rightarrow -i$ and phase transformation $b_q\rightarrow ib_q$, $b_q^+\rightarrow -ib_q^+$.
The Hamiltonian does not commute with the generators of each symmetry independently, but only with their product.
This is to be contrasted to the hermitian case, obtained by the replacement $i g_2\to g_2$ in Eq. \eqref{hamilton},
in which case
the sound velocity is\cite{giamarchi} $v_-=\sqrt{v^2-g_2^2}$.
In both cases, the system is stable for $|g_2|<v$, as we show below.
A distinct version of a non-hermitian LL was investigated in Ref. \cite{affleck2004}.

In the present work, we are interested in the time evolution 
of a LL following a sudden quantum quench. 
Initially, the system is
 prepared in the non-interacting ground state $|\phi_0\rangle$ (i.e. the vacuum for the $b$ bosons), 
and  time evolved with the non-hermitian Hamiltonian, Eq.  \eqref{hamilton} as $|\phi(t)\rangle=e^{-iHt}\,|\phi_0\rangle$.
~\footnote{Due to non-hermiticity, $\protect\langle\phi(t)|=\protect\langle\phi_0|e^{iH^+t}$.}.
In the hermitian realm, such systems were studied exhaustively\cite{cazalillaprl,karrasch,doraquench,iucci,dziarmagareview,polkovnikovrmp}.

In the fermionic realization of non-hermitian physics, the ensuing non-equilibrium
dynamics can be captured by the fermionic one-particle density matrix. The original fermion field
decomposes to right-going and a left-going parts\cite{giamarchi,nersesyan} as $\Psi(x)=e^{ik_F x}R(x) +e^{-ik_F x}L(x)$, 
therefore it is enough to investigate, for example, the correlator of the right-movers, defined as
\begin{gather}
G_r(x,t)\equiv \frac{\langle\phi(t)| R^+(x)R(0)|\phi(t)\rangle}{\langle\phi(t)|\phi(t)\rangle},
\label{green}
\end{gather}
describing excitations  around the right Fermi momentum, $k\approx k_F$.
The right-moving field,  $R(x)$, is expressed in terms of the
LL bosons as \cite{giamarchi}
$R(x)=\frac{\eta_r}{\sqrt{2\pi\alpha}}\exp\left(i\phi_r(x)\right)$,
where $\eta_r$ denotes the Klein factor,
and $\phi_r(x)=\sum_{q>0}\sqrt{2\pi/|q|L}e^{iqx-\alpha |q|/2}b_q+h.c.$ with $\alpha$ an ultraviolet regulator.
Due to the non-unitary time evolution\cite{carmichael,ashida},
it is compulsory to treat carefully the denominator arising in Eq. \eqref{green}.

\paragraph{The norm of the wavefunction.} To warm up, let us start by evaluating the denominator in Eq. \eqref{green}, which is $N(t)\equiv \langle\phi(t)|\phi(t)\rangle=\langle\phi_0|e^{iH^+t}e^{-iHt}|\phi_0\rangle
$, which would be 1 in the hermitian case.
Calculating $N(t)$ is accomplished  by realizing that the operators appearing in the Hamiltonian,
$K_0(q)=(b^+_qb_{q}+b_{-q}b^+_{-q})/{2}$, $K_+(q)=b^+_qb^+_{-q}$ and $K_-(q)=b_qb_{-q}$
 are the generators of  SU(1,1) Lie algebra~\footnote{The generators of the SU(1,1) Lie algebra satisfy $[K_+(q),K_-(q)]=-2K_0(q)$, $[K_0(q),K_\pm(q)]=\pm K_\pm(q)$, and the operators for distinct $q$'s commute with each other.}.
Exploiting a faithful $2\times 2$ matrix representation of the SU(1,1) generators ~\cite{solomon,gilmore}, the product of the time evolution operators is recast as
\begin{gather}
e^{iH^+t}e^{-iHt}=\prod_{q>0}e^{C_+(q,t) K_+(q)}e^{C_0(q,t) K_0(q)}e^{C_-(q,t) K_-(q)}.
\label{2timeevol}
\end{gather}
When taking its expectation value with the bosonic vacuum, the first and last exponentials containing $K_+(q)$ and $K_-(q)$ are Taylor expanded, and only the zeroth order term remains finite, all other terms containing powers of $b_qb_{-q}$ vanish when acting
on the vacuum. Then, the expectation value of Eq. \eqref{2timeevol} reduces to $\prod_{q>0}\exp(C_0(q,t)/2)$.
This is evaluated to yield
$N(t)=\prod_{q>0}\tilde v^2/(\tilde v^2-2g_2^2\sin^2(\omega_qt))$, 
which indeed gives one for $t=0$~\footnote{The steady state, $t\rightarrow\infty$ limit of the overlap is evaluated analytically in the thermodynamic limit as
$N(t\rightarrow\infty)=\left(\frac 12+\frac 12\frac{v_-}{\tilde v}\right)^{-L/\pi\alpha}$,
being valid for $v>|g_2|$.}.
Interestingly, the non-hermitian formulation remains valid only for $g_2<v$, similarly to the hermitian case, 
even though the renormalized sound velocity does not vanish for $v=|g_2|$.
\jav{The norm, $N(t)$} should always be non-negative. However, with increasing $g_2$, it first diverges and becomes negative for $|g_2|>v$.  
This happens because after the  quench, the time evolved
 wavefunction leaves the space of normalizable wavefunction, which is signaled by the total norm becoming negative\cite{graefe15,fishernieto}. 
 This behavior can be associated with the dynamical manifestation of the equilibrium instability  found in related systems\cite{ashida16}.

\paragraph{The numerator of the Green's function.} 
The two exponentials in the right-moving fields are merged using standard tricks\cite{delft} 
and the  time evolution of this operator is then evaluated using the identity
\begin{gather}
\langle\phi_0| e^{iH^+t} e^{-i(\phi_r(x)-\phi_r(0))} e^{-iHt}|\phi_0\rangle=\nonumber\\
=\langle\phi_0| e^{iH^+t} e^{-iHt} e^{iHt}  e^{-i(\phi_r(x)-\phi_r(0))} e^{-iHt}|\phi_0\rangle=\nonumber\\
\langle\phi_0| e^{iH^+t} e^{-iHt}  e^{-i(\phi_r(x,t)-\phi_r(0,t))} |\phi_0\rangle.
\label{timeevol}
\end{gather}
This allows us to formally define a pseudo-Heisenberg-type time evolution for the operators as
$b_q(t)=e^{iHt}b_qe^{-iHt}$, though this is {\it not} the physical Heisenberg time evolution as that would involve the
$e^{iH^+t}$ operator to the front.
The resulting pseudo-Heisenberg equation of motion is solved from $\partial_tb_q=i[H,b_q]$ and $\partial_tb^+_{-q}=i[H,b^+_{-q}]$,
which are {\it not} related to each other by hermitian conjugation.
This equation of motion is solved as
\
\begin{subequations}
\begin{gather}
b_q(t)=u_q(t)b_q+v_q(t) b^+_{-q}, \\
b^+_{-q}(t)=u^*_q(t) b^+_{-q}-v_q(t) b_q,
\end{gather}
\label{bogoliubovtimeevol}
\end{subequations}
\
and $|u_q(t)|^2+|v_q(t)|^2=1$~\footnote{Note the unconventional {"}+{"} sign as opposed to the conventional {"}-{"} for the hermitian case\cite{doraquench}.}, and the canonical commutation relation, $[b_q(t),b^+_q(t)]=1$ is preserved.
Finally, the time-dependent pseudo-Bogoliubov coefficients are
\begin{gather}
u_q(t)=\cos(\omega_qt)-\frac{iv}{\tilde v}\sin(\omega_qt),\hspace*{1mm}
v_q(t)=\frac{g_2}{\tilde v}\sin(\omega_qt),
\end{gather}
which are related to the hermitian Bogoliubov coefficients\cite{iucci} through the $g_2\rightarrow ig_2$ change.

The two time evolution operators, acting on the bra vector, are then rewritten using Eq. \eqref{2timeevol}
as
\begin{gather}
\langle\phi_0|e^{iH^+t} e^{-iHt}=\langle\phi_0| \prod_{q>0}e^{C_-(q,t) K_-(q)+\frac{C_0(q,t)}{2}}=\nonumber\\
=N(t)\langle\phi_0| \prod_{q>0} e^{C_-(q,t) K_-(q)}.
\label{leftpart}
\end{gather}
Therefore, the denominator appears also in the numerator and drops out from the final expression.
Using again the faithful representation of the SU(1,1) algebra, we get $C_-(q,t)=2u_q(t)v_q(t)/(|u_q(t)|^2-|v_q(t)|^2)$.

\paragraph{Supersonic modes. }
In order to calculate the vacuum expectation value, we normal order the pseudo-Heisenberg time evolved bosonic operators in the Green's function.
Using the pseudo-Heisenberg time evolution from Eq. \eqref{bogoliubovtimeevol}, we obtain 
\begin{gather}
e^{-i(\phi_r(x,t)-\phi_r(0,t))}=e^{-i\phi^+(x,t)}e^{-i\phi^-(x,t)}e^{c(x,t)},
\label{rightpart}
\end{gather}
where
$\phi^+(x,t)=\sum_{q>0}\sqrt{\frac{2\pi}{qL}}((e^{-iqx}-1)u^*_q(t)b^+_q+(e^{iqx}-1)v_q(t)b^+_{-q})$,
$\phi^-(x,t)=\sum_{q>0}\sqrt{\frac{2\pi}{qL}}((e^{iqx}-1)u_q(t)b_q-(e^{-iqx}-1)v_q(t)b_{-q})$,
$c(x,t)=\sum_{q>0}\frac{\pi}{qL}|e^{iqx}-1|^2(2|v^2_q(t)|-1)$.
In the hermitian case, the calculation would end here\cite{doraquench}, which contains all equilibrium and quench induced correlations, since $C_-(q,t)$ would be zero.
For the non-hermitian quench, 
by combining Eqs. \eqref{leftpart} and \eqref{rightpart}, the three exponentials are again Taylor expanded to calculate the required vacuum expectation value.
The $e^{-i\phi^-(x,t)}$ term gives one when acting on the vacuum.
The expansion of the $e^{C_-(q,t) K_-(q)}$ contains the same powers of $b_q$ and $b_{-q}$ due to the very definition of $K_-(q)$. Therefore, in order to have a non-zero expectation value, only those terms
contribute from the expansion of $e^{-i\phi^+(x,t)}$, which also contain the same powers of $b^+_q$ and $b^+_{-q}$.
This finally gives after some tedious algebra~\footnote{An identical expression applies to the left-movers as well.}
\begin{gather}
\label{greenq}
\frac{G_r(x,t)}{G^{0}_r(x)}=\exp\left(-\frac{8\pi}{L}\sum_{q>0}\frac{g_2^2\sin^2(qx/2)\sin^2(\omega_qt)}
{q(\tilde v^2-2g_2^2\sin^2(\omega_qt))}\right)
\end{gather}
where $G^{0}_r(x)={i}/({2\pi(x+i\alpha)})$ denotes the free fermion propagator, $L$ the system size.
This final result differs from the outcome of a hermitian quantum quench by the denominator in the exponent, but as we discuss below, it has profound consequences for the time evolution and light-cone structure.

The exponent of the Green's function is Taylor expanded in terms of $\sin^2(\omega_qt)$. Then, the various $q$ integrals are performed and the series is resumed, yielding
\begin{gather}
G_r(x,t)=G^{0}_r(x)\exp\left(\left(1-\frac{\tilde v}{v_-}\right)d(x,0)-\right.\nonumber\\
-2\left.\sum\limits_{n=1}^\infty\frac{\tilde v}{v_-}\left(\frac{-g_2^2}{v^2+\tilde vv_-}\right)^nd(x,nt)\right),
\label{greenx}
\end{gather}
where 
$d(x,t)=\frac 14\ln\left(\frac{(\alpha^2+(x-2\tilde v t)^2)(\alpha^2+(x+2\tilde vt)^2)}{(\alpha^2+4\tilde v^2t^2)^2}\right)$
using the $e^{-\alpha|q|}$ cutoff in Eq. \eqref{greenq}~\footnote{Note that other cutoff schemes yield distinct $d(x,t)$ functions, but its $x\gg \tilde 2vt$ and $x\ll 2\tilde vt$ behaviour remains unchanged.}.

For $2\tilde vt\gg x$, 
the Green's function becomes completely time-independent, similarly to the hermitian quench\cite{cazalillaprl}.
The characteristic non-integer power law decay of LL is observed as $G_r(x,t\rightarrow\infty)\sim |x|^{-\tilde v/v_-}$, 
and the exponent is smaller than for a hermitian quench\cite{iucci} with the same interaction strength $g_2$.
\jav{This establishes the LL universality also in the non-equilibrium and non-hermitian case.}

\begin{figure}[h!]
\includegraphics[width=7cm]{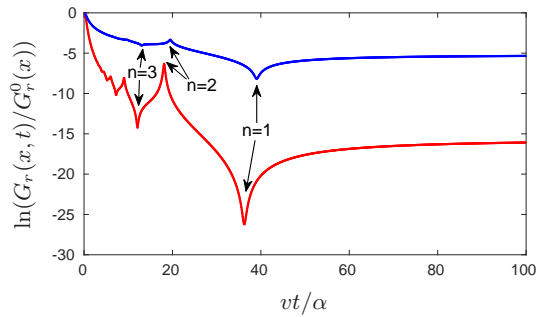}
\caption{The single particle density matrix is plotted from Eq. \eqref{greenx} for $g_2/v=0.8$ (blue) and 0.95 (red) line with $x=100\alpha$. The $n=1$ light-cone from the sum in $G_r(x,t)$ corresponds to the conventional light-cone with $2\tilde v$ velocity, 
while the first two supersonic features are denoted by $n=2$ and 3 with $4\tilde v$ and $6\tilde v$ velocity, respectively.}
\label{nhgreen}
\end{figure}

On the other hand, for $2\tilde vt\ll x$, supersonic modes that  propagate faster than the sound velocity
$\tilde v$, emerge and the corresponding velocities  are integer multiples of $2\tilde v$, even though $H$ itself is local\cite{liebrobinson}.
Its origin is traced back to the $e^{iH^+t} e^{-iHt}$ factor in Eq. \eqref{timeevol}. When merging
 the exponentials into a single one, a series of nested commutators arise from the Baker-Campbell-Hausdorff formula\cite{gilmore}
 due to $[H,H^+]\neq 0$, and the resulting exponent, interpreted as an effective Hamiltonian,  becomes
increasingly non-local and long range, therefore there is no obvious bound of the propagation speed of correlations \jav{in non-hermitian systems}. 
This parallels to the appearance of supersonic modes in hermitian long range systems\cite{eisert13}.
These are manifested in the denominator of Eq. \eqref{greenq}: upon expanding it in Taylor series in $\sin^2(\omega_qt)$,
the resulting expression involves terms that oscillate at frequencies $2\omega_q$, $4\omega_q$, $6\omega_q$\dots,
leading to the propagation velocities $2n\tilde v$. However, the  sharpness of the supersonic light-cones at $x=2n\tilde v t$
decreases with $n$ due to the $g_2^{2n}$ factor in Eq. \eqref{greenx}, as shown in Fig. \ref{nhgreen}.
By neglecting the $C_-(q,t)$ term, arising from $[H,H^+]\neq 0$, only a single conventional light-cone would appear.

We argue that 
supersonic modes appear generically during the time evolution of any correlator of local observables $\mathcal{O}_x$ \jav{in non-hermitian systems}.
Consider the correlation function $\chi(x,t)\equiv { \langle \phi(t)|\mathcal{O}_x\mathcal{O}_0|\phi(t)\rangle}/N(t)$ as
\begin{gather}
\chi(x,t)={\langle\phi_0| e^{iH^+t} e^{-iHt}   \mathcal{O}_x(t)\mathcal{O}_0(t)|\phi_0\rangle}/{N(t)},
\label{genericoperator}
\end{gather}
where $\mathcal{O}_x(t)=e^{iHt} \mathcal{O}_x e^{-iHt}$ is the pseudo-Heisenberg time evolved operator. Due to $e^{iH^+t} e^{-iHt}$, supersonic modes are 
expected from the argument below Eq. \eqref{greenx}.
Indeed, using Eq. \eqref{leftpart}, it is rewritten as
\begin{gather}
\chi(x,t)=\langle\phi_0|  e^{\sum_{q>0}C_-(q,t) K_-(q)}  \mathcal{O}_x(t)\mathcal{O}_0(t)|\phi_0\rangle.
\end{gather}
The $C_-(q,t)$ function contains $1/(\tilde v^2-2g_2^2\sin^2(\omega_qt))$, 
and when the expectation value is taken, this will inevitable alter the propagation velocity in $\chi(x,t)$ by
even integer multiples of $\tilde v$, similarly to the single particle density matrix in Eq. \eqref{greenx}.
Supersonic modes follow from the proper Heisenberg picture using the equation of motion method\cite{epaps}.
\jav{While supersonic modes were also seen in a non-hermitian non-interacting system\cite{ashida18}, our
results imply that these are expected on general ground in non-hermitian dynamics.}

\begin{figure}[h!]
\includegraphics[width=7cm]{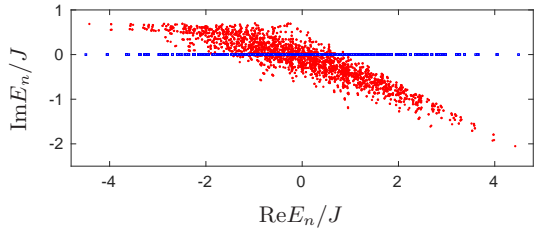}
\caption{The complex many-body energy spectrum, $E_n$ of Eq. \eqref{xxz} for $N=14$ at half filling is plotted
for $J_z=-0.3J$ (red dots) and the non-interacting, hermitian case with $J_z=0$ (blue squares) for comparison from exact diagonalization using
periodic boundary conditions.
During the time evolution with $e^{-itH}$, states with the largest imaginary part contribute the most,
which is the upper flat part of the spectrum. Out of these, states with increasing Re$E_n$ have less influence due to their decreasing overlap with the initial state. Due to the normalization in Eq. \eqref{genericoperator},
the imaginary offset of energies drops out from the expectation values.}
\label{spectrumpt}
\end{figure}

\paragraph{Numerics for lattice fermions.}
In order to test our results, we study a simple lattice model with imaginary interactions, which is not PT-symmetric, albeit 
the low energy part of its spectrum can be considered effectively real, while some higher modes develop significant imaginary parts, which would only
influence the long time dynamics.
The Hamiltonian is
\begin{gather}
H=\sum_{m=1}^N \frac{J+iJ_z}{2} \left(c^+_{m+1}c_m +\textmd{h.c.}\right)-i\frac{J_z\pi}{2} n_{m+1}n_m,
\label{xxz}
\end{gather}
where $c$'s are fermionic operators and $N$ the number of lattice sites, $n_m=c^+_mc_m$ and $J_z$ denotes the nearest neighbour interaction and the system is half filled.
Its low energy excitations are sound waves with sound velocity $\tilde v\approx J+(\pi^2/8-1)J_z^2/J$ after setting the lattice constant to one, which allows 
us to identify $g_2\sim -J_z/1.4$ for small $J_z$, while $v=J$.
Let us note, that the hermitian version ($iJ_z\to J_z$ in Eq.~\eqref{xxz}) is  Bethe-Ansatz solvable\cite{giamarchi}  with sound velocity  $v_-\approx J+(1-\pi^2/8)J_z^2/J$, 
in perfect agreement with the bosonization discussion following Eq.~\eqref{hamilton}.
The main merit of introducing $J_z$ into the hopping as well is that it eliminates the $g_4$ process which is only responsible for velocity 
renormalization\cite{giamarchi} but
does not induce non-integer power law decay of correlation functions, and makes the velocity real\cite{giamarchi} for the non-hermitian case.
Its energy spectrum at half filling obtained with exact diagonalization 
 is shown in Fig. \ref{spectrumpt}.

\begin{figure}[h!]
\includegraphics[width=8cm]{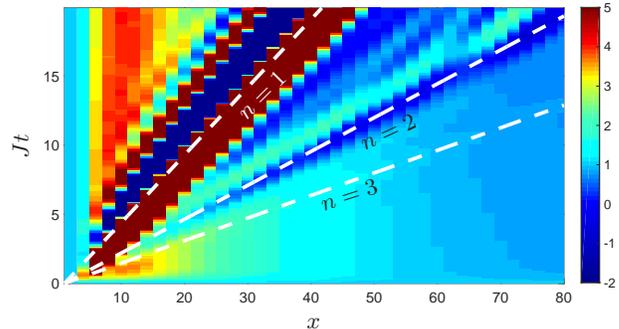}
\caption{Contour plot of the the density correlator, $\chi_{nn}(x,t)/\chi_{nn}(x,0)$, where denominator cancels 
the initial spatial correlation in the ground state, and all features result from the non-hermitian quench dynamics with $J_z/J=-0.3$.
The white dashed lines denote the $n=1$, 2 and 3 modes by using the sound velocity, $\tilde v=1.02J$ without any fitting. 
}
\label{nnsurf3d}
\end{figure}

We consider numerically a quench dynamics, when the system is prepared initially in the non-interacting, $J_z=0$ ground state of Hamiltonian, Eq.~\eqref{xxz}
as a Slater determinant. This is determined 
by the density matrix renormalization group~\cite{white1992} approach. Then,  we suddenly switch on $J_z$ and 
let the system evolve according to the Hamiltonian, Eq.~\eqref{xxz}. To study the quench dynamics we use the 
time evolving block decimation algorithm~\cite{Vidal2003} in the matrix product state representation. 
We have followed the time evolution of several 
physical quantities, such as the
single particle density matrix or the density correlator, and we found that all show signs of supersonic modes. However, there is a compromise: we have to keep the ration $|J_z/J|\lesssim 0.4 $, relatively small in order to retain the flat part of the spectrum as in Fig. \ref{spectrumpt} with constant imaginary part. 
On the other hand, the smallness of the interaction suppresses the higher order supersonic modes, as
evident from Fig. \ref{nhgreen}. In Fig. \ref{nnsurf3d}, we show the density correlator $\chi_{nn}(x,t)$, 
defined in Eq. \eqref{genericoperator} using $\mathcal{O}_x=n_x$ in a system with $N=201$ and 101 fermions. The
system is slightly away from half filling, which helps in killing the umklapp term\cite{giamarchi}. We checked that qualitatively similar results arise exactly at half filling with $N=200$ and 100 particles, though.

\jav{\paragraph{Experimental relevance.} 
Non-hermitian Hamiltonians can arise from a variety of different ways. It can arise from classical photonic waveguides, which only emulates the Schr\"odinger 
equation\cite{xiao}. In this case, supersonic modes are expected to occur
for arbitrary long times.
The non-hermitian time evolution can also follow from a conditional Lindblad-type dynamics\cite{ashida18,ashida16}, when the environment is continuously monitored in
order to maintain the condition of no quantum jump\cite{daley}.
The probability of having no quantum jump decreases steadily with time, rendering long time non-hermitian dynamics increasingly difficult to observe,
though not impossible\cite{avila}.
In this setting, Eq. \eqref{xxz} is realized by a 
dissipative lattice\cite{rauer} with one-body loss, i.e. superimposing a
weak resonant  optical  lattice (for the non-hermitian hopping) and also implementing background two-body loss\cite{daley} (for the imaginary interactions).
Finally, non-hermiticity could arise by adding the imaginary self energy (i.e. from finite lifetime quasiparticles) from diagrammatics\cite{kozii} to
 an originally Hermitian system, and treating this as an effective non-hermitian system.
 In this case, however, the temporal dynamics of the system would also require to go beyond the self-energy approximation and include vertex corrections as well, 
and the ensuing dynamics would probably feature a single light cone, as dictated by the Lieb-Robinson bound\cite{liebrobinson}.}

\paragraph{Summary.}
We studied many-body non-hermitian dynamics by a quantum quench in a PT-symmetric LL. The fermionic single particle density matrix reveals displays LL 
\emph{universality} in the long time limit.
For short times, in contrast to unitary evolution, the Lieb-Robinson bound is violated and
supersonic modes emerge due to non-unitary dynamics. They travel with velocities that are multiples of the conventional light-cone velocity.
\jav{We argue and demonstrate that } this emergent phenomena is characteristic to non-hermitian systems and arise from 
an effective long range Hamiltonian, governing the time
 evolution, although the physical Hamiltonian contains only short range, though non-hermitian terms.
Our analytical findings are benchmarked 
by the numerical study of a non-hermitian short range interacting lattice fermions.

\begin{acknowledgments}
We acknowledge a useful exchange of e-mails with Y. Ashida.
This research is supported by the National Research, Development and Innovation Office - NKFIH   within the Quantum Technology National Excellence Program (Project No.
      2017-1.2.1-NKP-2017-00001), K119442 and by
 a grant from the Simons Foundation.
This work was performed in part at Aspen Center for Physics, which is supported by National Science Foundation grant PHY-1607611.
\end{acknowledgments}

\bibliographystyle{apsrev}
\bibliography{wboson1,refgraph}

\newpage

\clearpage

\section{Supplementary material for "Quantum quench in PT-symmetric Luttinger liquid"}

\setcounter{equation}{0}
\renewcommand{\theequation}{S\arabic{equation}}

\setcounter{figure}{0}
\renewcommand{\thefigure}{S\arabic{figure}}

\section{Supersonic modes from the equation of motion}

The Hamiltonian in Eq. (1) in the main text is readily diagonalized\cite{garcia}, yielding a single excitation spectrum with a single velocity as
\begin{gather}
\omega_q=\tilde v |q| \textmd{  with } \tilde v=\sqrt{v^2+g_2^2}.
\label{spectrum}
\end{gather}
The Hamiltonian is quadratic in the bosons, therefore one could naively assume that the equation of motion method for the boson creation and annihilation operators would only feature
this velocity. This expectation, however, is not true in the non-hermitian case. The equation of motion for $b_q$ yields an infinite set of coupled differential equations, which contain multiples of the excitation spectrum in Eq. \eqref{spectrum}, in contrast to the hermitian case.

The proper Heisenberg equation of motion from a non-hermitian Hamiltonian for any operator reads as\cite{graefe2008} $\partial_tA=i(H^+A-AH)=i[\frac{H+H^+}{2},A]_--i[\frac{H-H^+}{2},A]_+$ due to the non-hermiticity, with $[\dots]_\pm$ being the anticommutator/commutator,
respectively.
We note that this should not be confused with the pseudo-Heisenberg equation of motion, discussed in Eqs. (6) in the main text.
Therefore, for the bosonic operators, the  Heisenberg equation of motion gives   
\begin{subequations}
\begin{gather}
\partial_t b_q=-iv|q|b_q+g_2|q|b^+_{-q}+2g_2|q|b^+_qb_qb^+_{-q},\\
\partial_t b^+_{-q}=iv|q|b^+_{-q}+g_2|q|b_{q}+2g_2|q|b^+_{-q}b_{-q}b_q
\end{gather}
\end{subequations}
which contain an anomalous third term on the r.h.s. due to non-hermiticity. In the hermitian case, only the first 2 terms on the r.h.s. are present, and we get a closed set of (i.e. two) 
coupled equations of motion, which
 are easily solved\cite{iucci}. These would only contain a single excitation spectrum. In the non-hermitian case, 
one has to study the equation of motion of the $b^+_qb_qb^+_{-q}$ and $b^+_{-q}b_{-q}b_q$ terms as well, which would generate another operators consisting of a large number of boson creation and annihilation
operators, namely an infinite set of coupled differential equations. By solving these, one could end up with higher harmonics of the  sound velocity in Eq. \eqref{spectrum}, which are responsible for the supersonic modes.

Another way to see the emergence of the supersonic modes in the time evolution of $b_q$ is to rewrite the proper Heisenberg time evolution  using the pseudo-Heisenberg time evolution as
\begin{gather}
e^{iH^+t}b_qe^{-iHt}=e^{iH^+t}e^{-iHt}e^{iHt}b_qe^{-iHt},
\end{gather}
where the $e^{iHt}b_qe^{-iHt}$ describes the pseudo-Heisenberg type time evolution and is solved in Eqs. (6) in the main text. This only contains a single velocity. However, as argued in the main text, by merging the
$e^{iH^+t}e^{-iHt}$ operators into a single exponential yields an infinitely long range effective Hamiltonian, which is responsible for the supersonic modes already at the level of simple bosonic operators.

To conclude, supersonic modes are expected whenever the $[H^+,H]\neq 0$ relation holds, irrespective of the details of the microscopic Hamiltonian $H$.

\end{document}